**Optimized Utilization of COMB3 Reactive Potentials in LAMMPS**

Robert Slapikas[1], Ismaila Dabo[1,2,3], and Susan B. Sinnott[1,2,4*]

[1)] Department of Materials Science and Engineering, The Pennsylvania State University, University Park, Pennsylvania 16801

[2)] Materials Research Institute, The Pennsylvania State University, University Park, Pennsylvania 16801

[3)] Institutes of Energy and the Environment, The Pennsylvania State University, University Park, Pennsylvania 16801

[4)] Department of Chemistry, The Pennsylvania State University, University Park, Pennsylvania 16801

***Corresponding author**

Email: sinnott@matse.psu.edu

An investigation to optimize the application of the third-generation charge optimized many-body (COMB3) interatomic potential and associated input parameters was carried out through the study of solid-liquid interactions in classical molecular dynamics (MD) simulations. The rates of these molecular interactions are understood though the wetting rates of water nano droplets on a bare copper (111) surface. Implementing the Langevin thermostat, the influence of simulation time step, the number of atoms in the system, the frequency at which charge equilibration is performed, and the temperature relaxation rate are all examined. The results indicate that time steps of 0.4 fs are possible when using longer relaxation times for the system temperature, which is almost double the typical time step used for reactive potentials. The use of the QEq charge equilibration allows for a fewer atomic layers to be used in the Cu slab. In addition, charge equilibrium schemes do not need to be performed every time step to ensure accurate charge transfer. Interestingly, the rate of wetting for the nanodroplets is dominantly dependent on the temperature relaxation time which are predicted to significantly change the viscosity of the water droplets. This work provides a pathway for optimizing simulations using the COMB3 reactive interatomic potential.

## I. Introduction

The recent push for renewable energy alternatives has placed an immense focus on understanding the fundamentals of electrochemical processes.[1,2] Computational modeling enables investigation of the thermodynamics and kinetics associated with processes that affect the atomic configurations and properties of materials in electrochemical systems.[3–5] One key modeling technique is molecular dynamics (MD) where Newton's second law of motion is integrated to predict individual atomic positions and momentums at future points in time.[3] Some of the most critical considerations in the application of classical MD simulations are the accuracy and computational cost, both of which depend on the interatomic potential (also called a force field) used to predict the atomic trajectories.[6]

Interatomic potential development is an art form that is not widely practiced because of the challenges associated with capturing the chemistry and physics needed to obtain accurate models while omitting negligible aspects to decrease the computational cost.[7] An analysis by Plimpton *et al.* reveals that every two years the computational cost of new potentials developed almost doubles and is roughly equivalent to Moore's Law for electronics.[7] Thus, plausible optimization of computational methods is perpetually a critical process that is further complicated by the fact that the choice of interatomic potentials for a given simulation is highly dependent upon the composition of the system being modeled.[6,7]

Some of the most significant and extensively observed electrochemical processes occur at solid-liquid interfaces.[1] At a molecular level surface science research aims to illustrate the fundamental thermodynamic and kinetic effects in this electrochemical phenomenon.[8–10] Dynamic wetting of a liquid droplet on a solid surface provides an experimental probe of fundamental solid-liquid properties such as surface tension of the liquid, viscosity of liquids, solid surface wettability, and the kinetic rates of wetting.[11–13] When a liquid droplet interacts with a solid surface, intermolecular interactions between the surfaces occur, bringing the system to equilibrium and causing partial wetting.[12,13]

Previous MD simulations of the solid-liquid interface to determine the molecular-kinetic theory for water droplets on a solid surface indicate that diffusion of the water droplet and its governing mechanism are complex.[12–15] However, in these simulations, the interactions between the water droplet and the solid surface were modeled using the Lennard-Jones 12-6 pair potential which is ideal for describing van der Waals interactions but is unable to capture electrostatic contributions to interatomic interactions at the solid-liquid interface.[16,17] Electrochemical or electrostatic interactions at the solid-liquid interface may be modeled using a variable charge bond order reactive potential such as ReaxFF or COMB3.[6,18,19] Recently, COMB3 was used in MD simulations to investigate solid-liquid interfaces by examining the self-diffusion of water nanodroplets on clean Cu and Pt surfaces, as well as Cu surfaces with oxygen and hydroxyl absorbates.[16–18] The results indicate that the water molecule interactions in the nanodroplets on the different surfaces change due to the chemical reactions of the unique solid-liquid molecular structures that are formed. By examining the molecular structures of these interfaces, it was determined that the rate of wetting is dependent on the strength and nature of the interactions between the water molecules and the surface.[16,17]

A significant amount of surface chemistry was found to be important at these particular interfaces and electrostatic effects are predicted to play a crucial role in the metal-liquid interface.[4,17] These simulations can be performed using the open source classical molecular dynamics code, Large-scale Atomic/Molecular Massively Parallel Simulator (LAMMPS), and the already incorporated COMB3 interatomic potential.[20] However, simulations of complex systems are extremely computationally expensive and take days to weeks to run. The objective of this work is to optimize the COMB3 potential within the LAMMPS software to reduce the computational cost, while maintaining physical and accurate simulation conditions. This work is accomplished through the study of molecular-kinetic theory as applied to solid-liquid interfaces and the optimization of the input parameters including the time step, rate of charge equilibration, relaxation time, and the number of atoms chosen to represent systems of interest. It is expected that this work will enable more efficient and realistic MD simulations using a reactive potential, particularly pertaining to solid-liquid interfaces.

## II.    Computational Details

In MD simulations the computational cost depends on the system that is modeled and the input parameters such as interatomic potential, type of ensemble, time step, and initial conditions. In previous MD simulations, an NVT ensemble and the dynamic charge interatomic potential COMB3 was used to model electrochemical phenomena at metal-water interfaces.[4,16,17] Consequently, the input parameters that will be examined to increase the speed of the simulation are the time step ($\delta t$), the number of atoms in the simulation, the rate of charge equilibration (QEq), and the rate at which the temperature of the system is relaxed in Brownian dynamics ($\tau_{damp}$).[21–23]

All dynamic wetting MD simulations are performed using the newly parameterized Cu-O-H potentials developed by Anthony et al.[17,24] In the simulations, a water droplet containing 576 $H_2O$ molecules is considered that is interacting with 142 Å × 142 Å Cu (111) slabs of different thickness, ranging from three atomic layers to nine atomic layers. The water nanodroplet is first obtained by performing NPT MD simulations of ice-$I_h$ at 300 K for 200. The system is controlled with two different NVT thermostats: a Langevin thermostat with a simulation input temperature of 370 K is used for the water to obtain an output temperature of 300 K, and a Nose-Hoover thermostat with an input temperature of 300 K is used for the Cu slabs to obtain an output temperature of 300 K.[21,22,25,26] In each of the Cu slabs the bottom layer of atoms is held rigid. Time steps of 0.1, 0.2, 0.25, and 0.4 fs were used, and a range of relaxation rates were tested from 0.002 to 1.0 ps. A single QEq charge equilibration scheme is performed for all the atoms in the simulations at rates of every 1, 5, 10, 25, 50, and 100 time step iterations.[23,27]

The simulations are carried out using periodic boundary conditions where the Coulombic energy of the system is calculated through the summation of ion pairs to allow charge transfer between atoms.[18] In COMB3 the Wolf summation method is used for Coulombic electrostatic interactions and a set cutoff range of 11.1 Å is used.[18,28] Which requires the simulation box size to be larger than 11.1 Å in all directions to ensuring that none of the atoms interact with themselves. Spreading rates are determined by calculating the radii of the absorbed water molecules on the different Cu slabs.

### III. Results and Discussion

The most obvious parameter to change in order to speed up a computational simulation is the time step. In MD simulations, a typical time step for variable charge bond order potentials is 0.25 fs.[4,7] However, due to a change in other input parameters such as $\tau_{damp}$, it is seen that the time step used in the simulations can be increased and longer simulation times can be achieved for the same number of time steps without affecting the spreading rates significantly (Figure 1). Contrary, to prior observations, a larger time step of 0.4 fs which has previously been associated with unstable simulations can be used when a lower $\tau_{damp}$ parameter is used.[4,7]

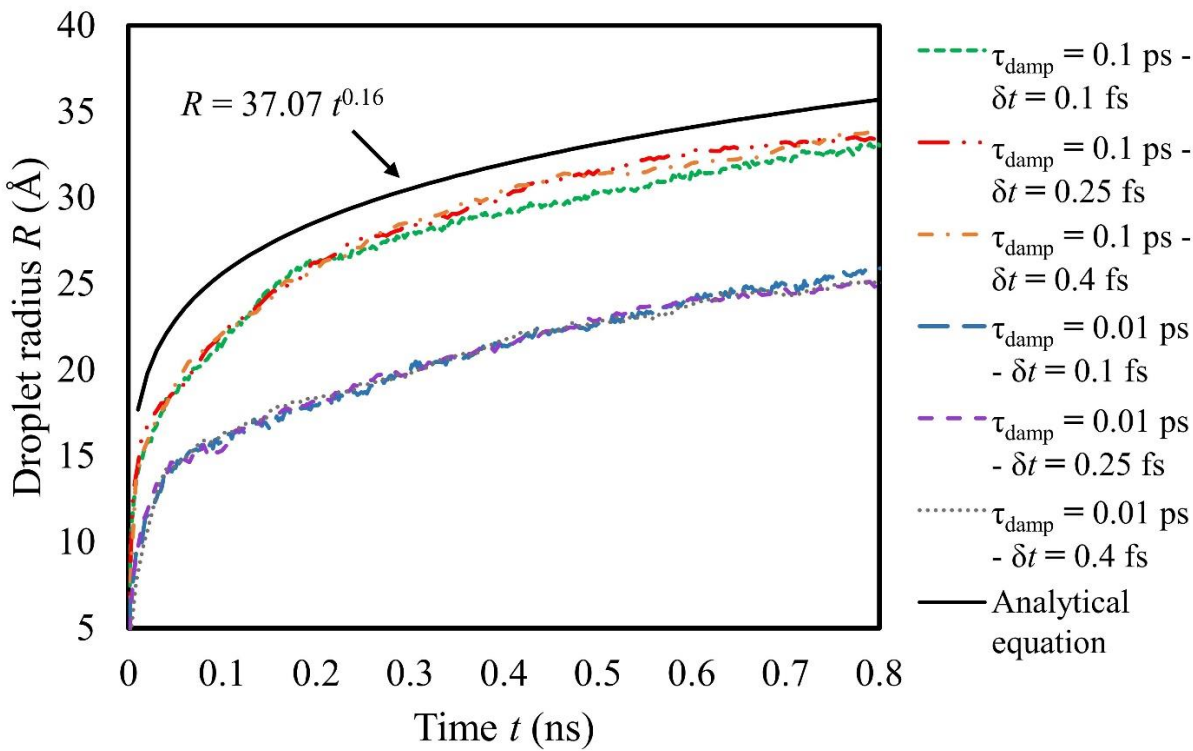

**FIG. 1**. Water nanodroplet spreading rate for different time steps ($\delta t$) in femtoseconds (fs) and different $\tau_{damp}$ rates on a Cu surface with 3 Cu layers as a function of simulation time. An analytical equation is used as a reference point for water molecules spreading on a Cu (111) surface.[16]

Another important consideration in running MD simulations is the size of the simulation and optimizing it to observe the desired physical processes. This is particularly relevant for the COMB3 interatomic potential which has a Coulombic cutoff distance of 11.1 Å which means the physical size of the simulation needs to be larger than this when periodic boundary conditions are enforced so the atom is not allowed to interact with itself.[18] However, by increasing the number of atoms in the simulation, the computational cost increases due to the higher number of calculations. Also, because Coulombic interactions are the driving force for the rate of wetting of the water nanodroplets, it is crucial to determine the way in which the number of Cu atomic layers in the substrate impact the speed of wetting. Since in previous work the substrate was only three atomic layers thick and below the 11.1 Å cutoff distance, the effects of the substrate thickness were considered by running the same MD simulations with the same conditions but varying the number of atomic layers from three to nine.[16,17] From these simulations (Figure 2), it was determined that the wetting rate of the nanodroplet remains constant and does not depend on the number of atomic layers. Thus, decreasing the number of atomic layers decreases simulation time by having fewer atoms to integrate over but maintains the same kinetic process of surface diffusion. While the most realistic charge neutrality of the system is not captured due to size and cost of the simulation, the error is small enough to be considered negligible. It is determined that this is an effect of using the electronegativity equilibration scheme QEq which enforces charge neutrality.[4,23] It is seen that for all the different layers, the Coulombic energy convergence occurs at the same rate, and closely reflects the spreading rate of the nanodroplet on the Cu surface.

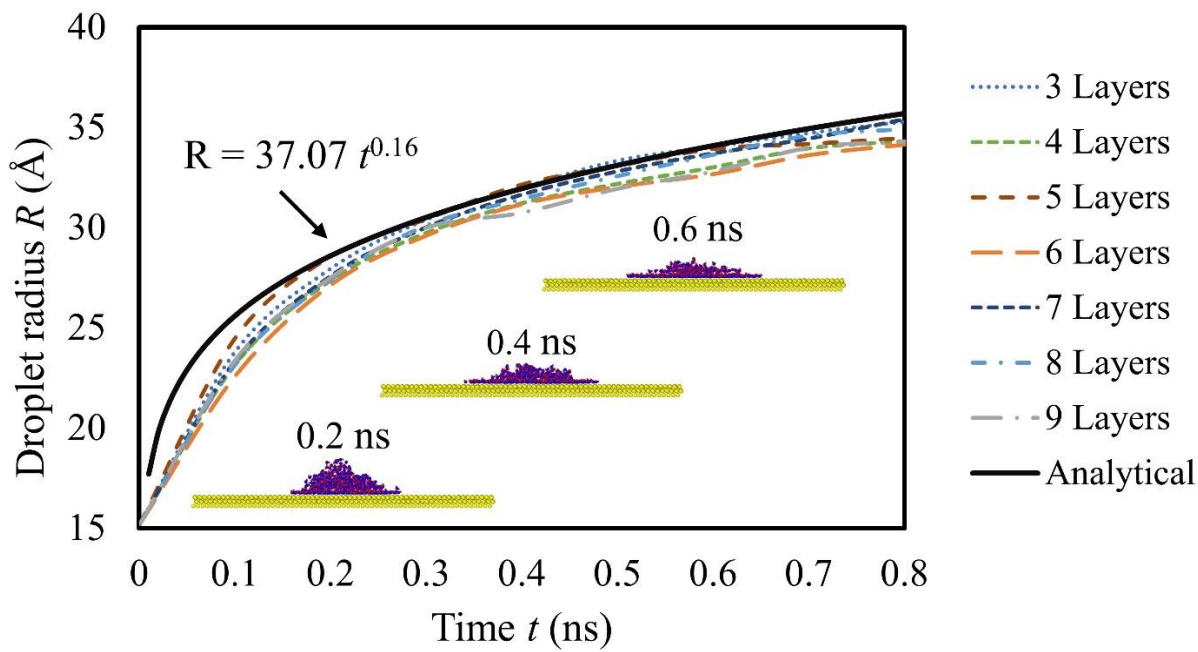

**FIG. 2**. Spreading rate of water nanodroplet for different number of Cu layers as a function of simulation time. (Insets) Snapshots of the simulation for 3 Cu layers. An analytical equation is used as a reference point for water molecules spreading on a Cu (111) surface.[16]

In the MD simulations performed for this work, two NVT ensembles are used to examine the solid-liquid interfaces, and the temperature of the water nano droplet is held constant using the Langevin thermostat.[21,22] By using this thermostat Newton's equations of motion are altered, and the force on an atom is described with Equation 1 where $F_c$ is the conservative force from the interatomic potential, $F_f$ is the frictional force given by Equation 2, and $F_r$ is the random force (white noise) whose time autocorrelation is given by Equation 3.[21,22] In these equations $m$ is the mass of an atom, $v$ is the velocity of the particle, $T$ is the temperature of the system, $k_B$ is the Boltzmann constant, and $\delta t$ is the time step.

$$F = F_c + F_f + F_r \tag{1}$$

$$F_f = -\frac{m}{\tau_{\text{damp}}} v \tag{2}$$

$$\langle F_r(t) \cdot F_r(t') \rangle \propto \frac{m k_B T}{\tau_{\text{damp}}} \delta(t-t') \tag{3}$$

By adjusting $\tau_{\text{damp}}$, it is seen that it is loosely related to the viscosity of a liquid through an inverse relationship between the self-diffusion coefficient and the viscosity of a liquid described by the Stokes-Einstein equation.[29] Therefore, a small $\tau_{\text{damp}}$ will increase the viscosity of a liquid and decrease the rate of wetting. This is depicted in Table I where the self-diffusion rates of the nanodroplets are listed and obtained from the mean squared displacement (MSD) graph (Figure 3) where the linear portions of each graph are taken as the self-diffusion coefficient of the Stokes-Einstein equation.[17,29] However, if the dampening parameter $\tau_{\text{damp}}$ is too large, the system will never reach thermodynamic equilibrium and the temperature of the system will drop to 0 K. Conversely, a small $\tau_{\text{damp}}$ causes the average temperature of the simulation to increase which causes the viscosity of the liquid to increase, a phenomenon that does not occur in water.[13,30] It is also observed that $\tau_{\text{damp}}$ affects the rate at which Coulombic energy convergences occur (Figure 4). In all three simulations, the same physical responses are observed at different rates when $\tau_{\text{damp}}$ is adjusted, and thermodynamic equilibrium is reached. Thus, by increasing $\tau_{\text{damp}}$ to the point right before thermodynamic equilibrium can no longer be reached, the kinetic reaction rates increase in the system and maintain the accuracy of realistic models. This in turn causes the computational cost of the simulations to decrease.

**TABLE I.** Showing effects of $\tau_{damp}$ on temperature and self-diffusion rates of the water nano droplets.

| $\tau_{damp}$ (ps) | Avg. Water Temp. (K) | Self-diffusion ($\text{Å}^2\text{ ns}^{-1}$) |
|---|---|---|
| 1.0 | 0 | N/A |
| 0.1 | 333 | 448 |
| 0.01 | 367 | 74 |
| 0.002 | 370 | 17 |

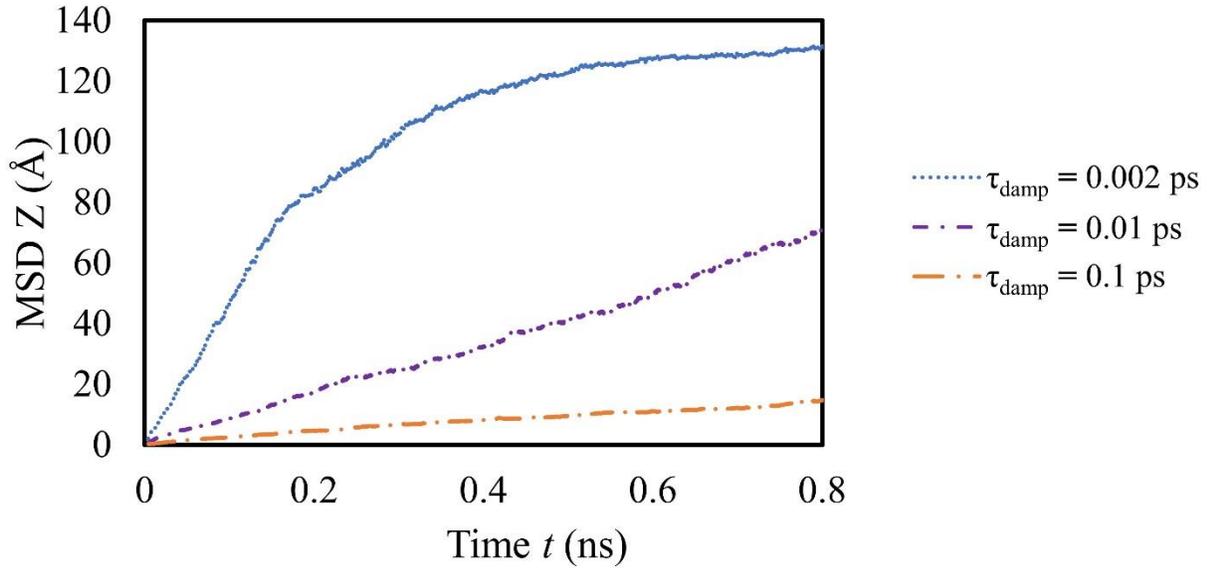

**FIG. 3.** MSD of water nanodroplet for different values of $\tau_{damp}$ in the in-plane direction of the Cu surface with 3 Cu layers as a function of simulation time.

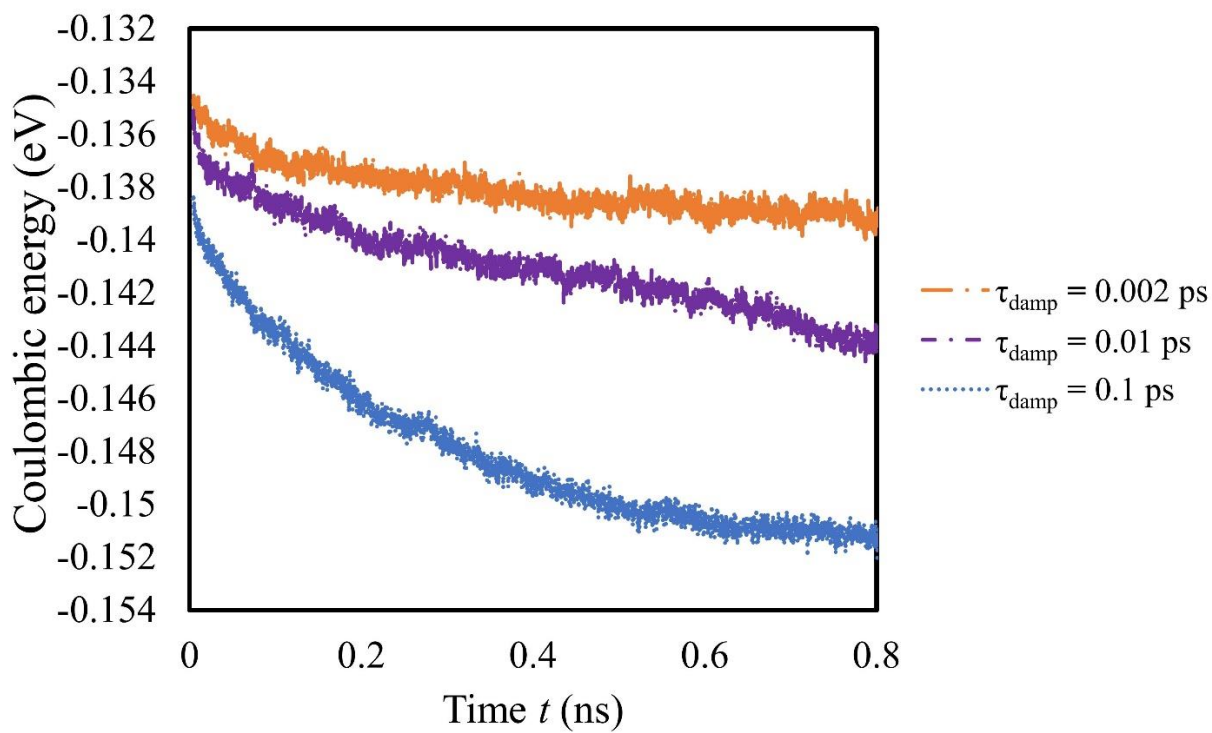

**FIG. 4.** Coulombic energy as a function of simulation time for water nanodroplet spreading on Cu surface with 3 Cu layers as a function of simulation time.

Another input parameter that was considered to optimize COMB3 potential was the frequency of charge equilibration for the system which is an important parameter for the accurate simulation of dynamical processes.[4,7] Charge equilibration is a complex calculation, since it is an iterative process dependent on an atoms local environment, that should be performed at every time step to obtain the most realistic physical event, making it a rigorous process to implement.[4,6,23] In this work, decreasing the simulation time for Coulombic charge convergence to occur was evaluated by performing charge equilibration at different time step iterations. However, when comparing different charge equilibration rates of every time step, every 10th time step, and every 50th time step, it is observed in Figure 5 that the amount of simulation time it takes for the Coulombic charge to converge does not change. Therefore, it was determined that the best way to decrease computational cost is to reduce the charge equilibration frequency while maintaining the accuracy of the physical processes. In these simulations, it was determined that performing charge equilibration every 50 steps vs every time step causes the computational burden to decrease by more than five times while still providing accurate spreading rates. It is also observed from Figures 5 and 6 that the $\tau_{damp}$ parameter is still more influential in the kinetic rates of the simulations since a large $\tau_{damp}$ achieves a quicker rate of wetting of the nano droplet and a quicker rate of Coulombic charge convergence.

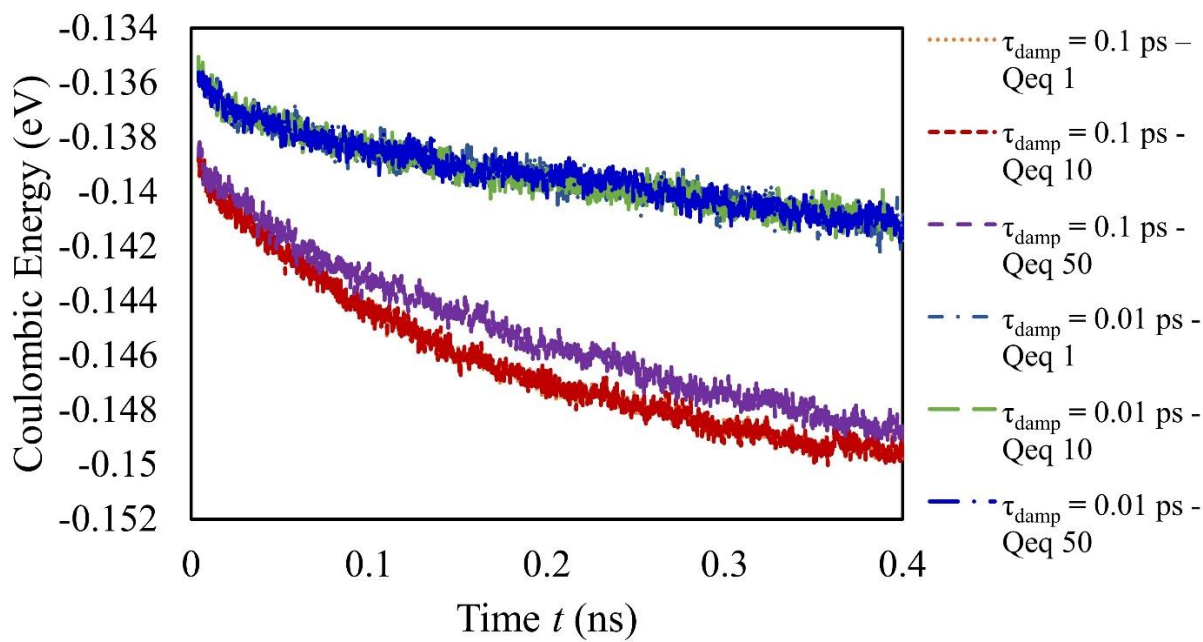

**FIG. 5.** Water nanodroplet spreading Coulombic energy as a function of simulation time for different charge equilibration per number of MD time steps (QEq) and different $\tau_{damp}$ rates on a Cu surface with 3 Cu layers as a function of simulation time.

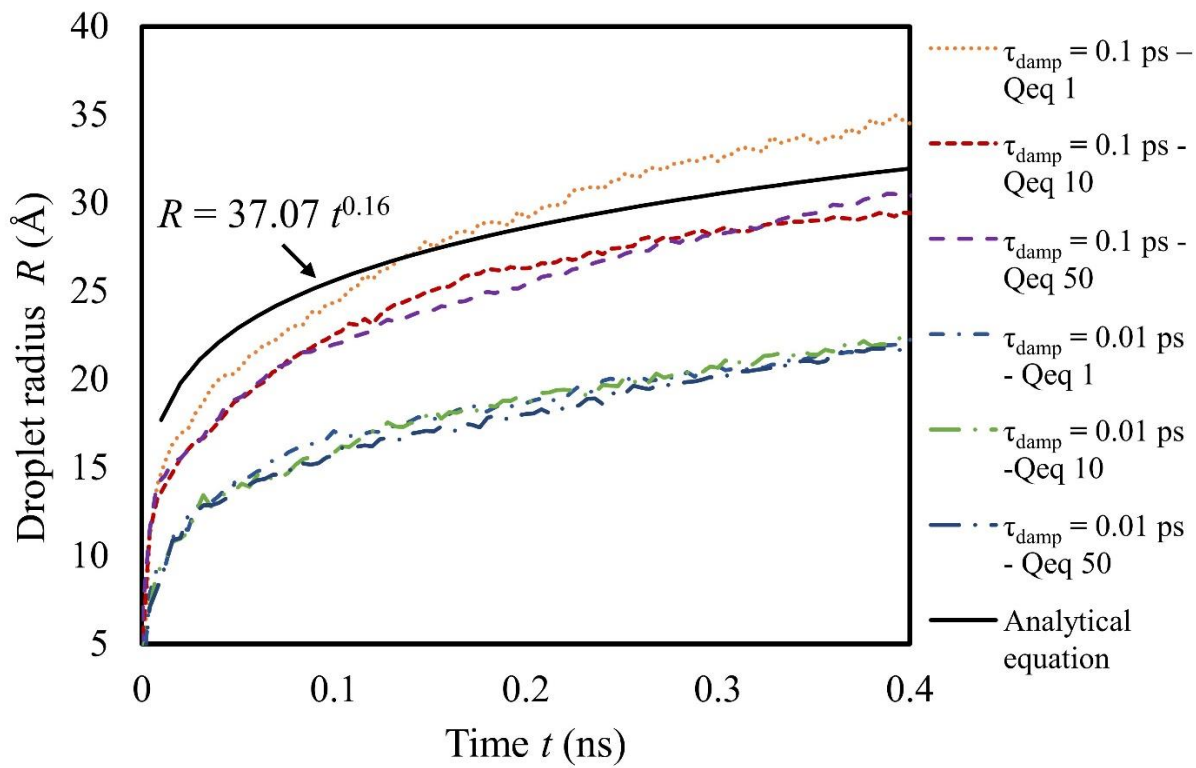

**FIG. 6.** Water nanodroplet spreading rate for different charge equilibration per number of MD time steps (QEq) and different $\tau_{damp}$ rates on a Cu surface with 3 Cu layers as a function of simulation time. An analytical equation is used as a reference point for water molecules spreading on a Cu (111) surface.[16]

## IV. Conclusions

We have used the COMB3 interatomic potential to model solid-liquid interactions, focusing on the spreading rates of water nanodroplets on bare Cu (111) slabs. To reduce the computational complexity of COMB3 simulations, we have carried out a systematic analysis of the different numerical parameters that affect the accuracy and efficiency of the time integration.

With the implementation of the QEq charge equilibration scheme in COMB3, and the system being driven to charge neutrality, the number of atomic layers in a substrate surface can be reduced in the direction out of plane to the surface to reduce the total number of atoms in the simulation. It was also determined that the charge equilibration calculations of the system does not need to occur every time step in order to achieve the same spreading rates. Lastly, changing how fast the temperature of the system is relaxed can either increase or decrease the rate at which the kinetic reactions occur and is considered to be the most influential input parameter. With this information, the scientific community can optimize the COMB3 potential while still maintaining realistic results to further the understanding of electrochemical processes at the atomic level.

**ACKNOWLEDGMENTS**

The authors acknowledge helpful discussions with Andrew C Antony and Tao Liang on their previous work with the COMB3 potential and the associated parameterization. We acknowledge financial support from the U.S. Department of Energy, Office of Science, Basic Energy Sciences, CPIMS Program, under Award No. DE-SC0018646. Computer simulations for this work were performed on the Pennsylvania State University's Institute for CyberScience Advanced CyberInfrastructure (ICS-ACI).

**AIP PUBLISHING DATA SHARING POLICY**

The data that support the findings of this study are available from the corresponding author upon reasonable request.

**REFERENCES**

[1] H. Li, J. Gazzarri, K. Tsay, S. Wu, H. Wang, J. Zhang, S. Wessel, R. Abouatallah, N. Joos, and

J. Schrooten, Electrochim. Acta **55**, 5823 (2010).

[2] T. Yoshida and K. Kojima, Electrochem. Soc. Interface **24**, 45 (2015).

[3] M.P. Allen and D.J. Tidesley, *Computer Simulation of Liquids / Michael P. Allen and Dominic J. Tildesley. - Version Details - Trove* (2017).

[4] T. Liang, A.C. Antony, S.A. Akhade, M.J. Janik, and S.B. Sinnott, J. Phys. Chem. A **122**, 631 (2018).

[5] M. Patra and M. Karttunen, J. Comput. Chem. **25**, 678 (2004).

[6] J.A. Harrison, J.D. Schall, S. Maskey, P.T. Mikulski, M.T. Knippenberg, and B.H. Morrow, Appl. Phys. Rev. **5**, 031104 (2018).

[7] S.J. Plimpton and A.P. Thompson, MRS Bull. **37**, 513 (2012).

[8] A. Verdaguer, G.M. Sacha, H. Bluhm, and M. Salmeron, Chem. Rev. **106**, 1478 (2006).

[9] M.A. Henderson, Surf. Sci. Rep. **46**, 1 (2002).

[10] A. Hodgson and S. Haq, Surf. Sci. Rep. **64**, 381 (2009).

[11] R.G. Cox, J. Fluid Mech. **168**, 169 (1986).

[12] M.J. de Ruijter, T.D. Blake, and J. De Coninck, Langmuir **15**, 7836 (1999).

[13] L. Chen and E. Bonaccurso, Phys. Rev. E - Stat. Nonlinear, Soft Matter Phys. **90**, 022401 (2014).

[14] N. Sedighi, S. Murad, and S.K. Aggarwal, Fluid Dyn. Res. **43**, (2011).

[15] C. Wu, T. Qian, and P. Sheng, J. Phys. Condens. Matter **22**, (2010).

[16] A.C. Antony, T. Liang, and S.B. Sinnott, (2018).

[17] A.C. Antony, T. Liang, S.A. Akhade, M.J. Janik, S.R. Phillpot, and S.B. Sinnott, Langmuir **32**, 8061 (2016).

[18] T. Liang, T.R. Shan, Y.T. Cheng, B.D. Devine, M. Noordhoek, Y. Li, Z. Lu, S.R. Phillpot, and S.B. Sinnott, Mater. Sci. Eng. R Reports **74**, 255 (2013).

[19] A.C.T. Van Duin, S. Dasgupta, F. Lorant, and W.A. Goddard, J. Phys. Chem. A **105**, 9396 (2001).

[20] S. Plimpton, *Fast Parallel Algorithms for Short-Range Molecular Dynamics* (1995).

[21] B. DÜNWEG and W. PAUL, Int. J. Mod. Phys. C **02**, 817 (1991).

[22] T. Schneider and E. Stoll, Phys. Rev. B **17**, 1302 (1978).

[23] A.K. Rappé and W.A. Goddard, J. Phys. Chem. **95**, 3358 (1991).

[24] A.C. Antony, S.A. Akhade, Z. Lu, T. Liang, M.J. Janik, S.R. Phillpot, and S.B. Sinnott, J. Phys. Condens. Matter **29**, (2017).

[25] S. Nosé, J. Chem. Phys. **81**, 511 (1984).

[26] W.G. Hoover, Phys. Rev. A **31**, 1695 (1985).

[27] S.W. Rick, S.J. Stuart, and B.J. Berne, *Dynamical Fluctuating Charge Force Fields: Application to Liquid Water* (n.d.).

[28] D. Wolf, P. Keblinski, S.R. Phillpot, and J. Eggebrecht, **8254**, (2012).

[29] S. Chandrasekhar, Rev. Mod. Phys. **15**, 1 (1943).

[30] L. Korson, W. Drost-Hansen, and F.J. Millero, J. Phys. Chem. **73**, 34 (1969).